\documentclass[10pt,pra,twocolumn,floatfix,showpacs]{revtex4}
\usepackage{amsfonts}
\usepackage{amssymb}
\usepackage{amsmath}
\usepackage[dvips]{graphicx}

\begin{document}

\title{Avoiding disentanglement of multipartite entangled optical beams with
a correlated noisy channel}
\author{Xiaowei Deng,$^{1,2}$ Caixing Tian,$^{1,2}$ Xiaolong Su$^{1,2}$$%
^{\ast }$ and Changde Xie$^{1,2}$} 
\affiliation{$^{1}$State Key Laboratory of Quantum Optics and Quantum Optics Devices,\\
Institute of Opto-Electronics, Shanxi University, Taiyuan 030006, China \\
$^{2}$Collaborative Innovation Center of Extreme Optics, Shanxi University,
Taiyuan 030006, China \\
$^*$Corresponding author, e-mail: suxl@sxu.edu.cn}

\begin{abstract}
A quantum communication network can be constructed by
distributing a multipartite entangled state to space-separated nodes.
Entangled optical beams with highest flying speed and measurable brightness
can be used as carriers to convey information in quantum communication
networks. Losses and noises existing in real communication channels will
reduce or even totally destroy entanglement. The phenomenon of
disentanglement will result in the complete failure of quantum
communication. Here, we present the experimental demonstrations on the
disentanglement and the entanglement revival of tripartite entangled optical
beams used in a quantum network. We discover that symmetric tripartite
entangled optical beams are robust in pure lossy but noiseless channels.
While in a noisy channel the excess noise will lead to the disentanglement
and the destroyed entanglement can be revived by the use of a correlated
noisy channel (non-Markovian environment). The presented results provide
useful technical references for establishing quantum networks.
\end{abstract}

\maketitle

\section*{INTRODUCTION}

Quantum entanglement is a fundamental resource in quantum information tasks 
\cite{Nielson}. Considerable progress has been made in quantum information
processing with entangled optical beams because the manipulation and
measurement of the quadrature amplitudes of optical field are familiar in
classical communication and processing technologies \cite%
{Lampingk,Braunstein,Weedbrook}. The used quantum variables, amplitude and
phase quadratures, are just the analogies of position and momentum of a
particle \cite{Lampingk}. Especially, the multipartite entangled state can
be used to complete one-way quantum computation \cite%
{Raussendorf2001,Walther2005,Ukai2011,Su} and to construct quantum
communication networks, such as quantum teleportation network \cite%
{Loock,Hide}, controlled dense coding quantum communication \cite{Jing}, and
wavelength-multiplexed quantum network with ultrafast frequency comb \cite%
{Jona}. In quantum communication networks, quantum states carrying
information are transmitted between space-separated nodes through quantum
channels, while losses and noises in channels will unavoidably lead to
decoherence of quantum states.

Decoherence, which is often caused by the interaction between system and the
environment, is a main factor limiting the development of the quantum
information technology. In quantum communication, the distributed
entanglement will decrease because of the unavoidable decoherence in the
quantum channel. In this case, entanglement purification, which is a way to
distill highly entangled states from less entangled ones, is a necessary
step to overcome decoherence \cite{Pan,Duan,Hage,Dong}. Furthermore, it has
been shown that decoherence will lead to entanglement sudden death (ESD) 
\cite{Yu,Almeida}, where two entangled qubits become completely disentangled
in a finite-time under the influence of vacuum. ESD for multipartite
entangled states has also been discussed theoretically \cite{Lopez,Aolita}.
Various methods to recover the bipartite entanglement after ESD occurred
have been proposed and demonstrated during past several years, such as the
non-Markovian environment \cite{Xu}, weak measurement \cite{Kim}, feedback 
\cite{Yamamoto}, \textit{et al.} The fidelity of quantum teleportation
directly depends on the entanglement degree of utilized quantum resource. If
disentanglement occurs in a quantum teleportation network, the fidelity will
never exceed its classical limit and thus the quantum communication will
fail. Thus, it is necessary to investigate the physical conditions of
reducing and destroying multipartite entanglement in quantum channels and
explore the feasible schemes of avoiding disentanglement.

Generally, there are two types of quantum channels, the lossy channel and
the noisy channel. In a lossy but noiseless (without excess noise) quantum
channel, the noise induced by loss is nothing but the vacuum noise
(corresponding to a zero-temperature environment) \cite{Weedbrook}. In a
noisy channel, the excess noise higher than the vacuum noise exists \cite%
{Weedbrook}. It has been shown that the bipartite Gaussian entangled optical
beams can be robust against loss \cite{Barbosa1}, while the three-color
entanglement among three asymmetric optical modes can be fragile against
loss, which means that the disentanglement is observed in a lossy channel,
at some situation \cite{Coelho}. We theoretically analyze the physical
condition of the robustness for tripartite entangled optical beams over
lossy channels, and experimentally demonstrate that the symmetric tripartite
entangled state is robust against loss in quantum channels.

The excess noise in a communication channel is another main factor limiting
the transmission of information, for example, it will decrease the secure
transmission distance of quantum key distribution \cite{Weedbrook}. The
noises in today's communication systems exhibit correlations in time and
space, thus it will be relevant to consider channels with correlated noise
(non-Markovian environment) \cite{Kret,Corney,Andersen}. A correlated noisy
channel has been used to complete Gaussian error correction \cite{Andersen}
and to protect squeezing in quantum communication with squeezed state over a
noisy channel \cite{Deng}. We study the entanglement property of the
tripartite entangled optical beams over a noisy quantum channel, in which
the disentanglement is observed. By applying an ancillary optical beam and
establishing a correlated noisy channel, we successfully avoid
disentanglement among the tripartite disentangled optical beams.
\begin{figure}[tbp]
\begin{center}
\includegraphics[width=90mm]{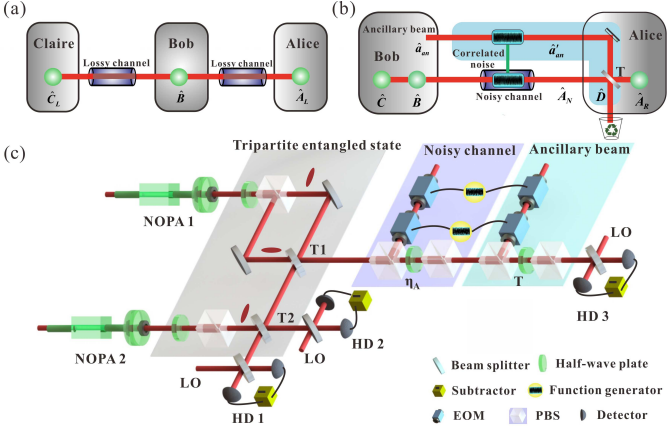}
\end{center}
\caption{\textbf{Schematic of principle and experimental set-up.} (\textbf{a}%
) Two modes ($\hat{A}$ and $\hat{C}$)\ of a tripartite entangled state are
distributed over two lossy quantum channels to Alice and Claire,
respectively. (\textbf{b}) One mode of the tripartite entangled state is
distributed over a noisy channel, where disentanglement is observed among
optical modes $\hat{A}_{N},\hat{B}$ and $\hat{C}$. The entanglement revival
operation is implemented by coupling an ancillary beam ($\hat{a}%
_{an}^{\prime }$) who has correlated noise with the environment with the
transmitted mode $\hat{A}_{N}$ on a beam-splitter with transmission
efficiency of $T$,\ thus the entanglement among modes $\hat{A}_{R}$, $\hat{B}
$\ and $\hat{C}$\ is revived. (c) The schematic of experimental set-up for
distributing a mode of the tripartite entangled optical beams over a noisy
channel and the entanglement revival. T1 and T2 are the beam-splitters used
to generate the GHZ entangled state.\ $\protect\eta _{A}$\ and $T$\ are the
transmission efficiencies of the noisy channel and the revival
beam-splitter, respectively. HD$1-3$, homodyne detectors. LO, the local
oscillator.}
\end{figure}

\section*{Experimental scheme}

The quantum state used in the experiment is a continuous variable
Greenberger-Horne-Zeilinger (GHZ) tripartite entangled state \cite{Hide,Jing}%
, which is prepared deterministically. The correlation variances between the
amplitude (position) and phase (momentum) quadratures of the tripartite
entangled state are expressed\emph{\ }by $\Delta ^{2}\left( \hat{x}_{A}-\hat{%
x}_{B}\right) =\Delta ^{2}\left( \hat{x}_{A}-\hat{x}_{C}\right) =\Delta
^{2}\left( \hat{x}_{B}-\hat{x}_{C}\right) =2e^{-2r}$ and $\Delta ^{2}\left( 
\hat{p}_{A}+\hat{p}_{B}+\hat{p}_{C}\right) =3e^{-2r}$, respectively, where
the subscripts correspond to different optical modes ($\hat{A}$, $\hat{B}$\
and\emph{\ }$\hat{C}$) and $r$\ is the squeezing parameter ($r=0$\ and $%
r=+\infty $\ correspond to no squeezing and the ideally perfect squeezing,
respectively). Obviously, in the ideal case with infinite squeezing ($%
r\rightarrow \infty $), these correlation variances will vanish and the
better the squeezing, the smaller the noise terms.

When two optical modes ($\hat{A}$\ and $\hat{C}$) of a tripartite entangled
state are distributed by Bob (who retains mode $\hat{B}$) to two nodes
(Alice and Claire) over two lossy channels, a quantum network with three
users Alice, Bob and Claire is established [Fig. 1(a)]. After the
transmission of optical modes $\hat{A}$\ and $\hat{C}$\ over two lossy
channels, the output modes are given by $\hat{A}_{L}=\sqrt{\eta _{A}}\hat{A}+%
\sqrt{1-\eta _{A}}\hat{\upsilon}_{A}$\ and $\hat{C}_{L}=\sqrt{\eta _{C}}\hat{%
C}+\sqrt{1-\eta _{C}}\hat{\upsilon}_{C}$, where $\eta _{A(C)}$\ and $\hat{%
\upsilon}_{A(C)}$\ represent the transmission efficiency of the quantum
channel and the vacuum state induced by loss into the quantum channel,
respectively. When $\eta _{A}\neq 1,\eta _{C}=1,$\ it corresponds to the
situation that the optical mode $\hat{A}$\ is distributed over a lossy
channel to another node while modes $\hat{B}$\ and $\hat{C}$\ are maintained
in a node. While when mode $\hat{A}$\ is distributed over a noisy channel
[Fig. 1(b)], the transmitted mode is expressed by%
\begin{equation}
\hat{A}_{N}=\sqrt{\eta _{A}}\hat{A}+\sqrt{1-\eta _{A}}\hat{\upsilon}_{A}+%
\sqrt{\left( 1-\eta _{A}\right) g_{a}}\hat{N},
\end{equation}%
where $\hat{N}$ and $g_{a}$ represent the Gaussian noise in the channel and
the magnitude of noise, respectively. The excess noise on the transmitted
mode will possibly lead to the disentanglement of the tripartite entangled
state. In order to avoid disentanglement of the tripartite entangled state,
an ancillary beam with correlated noise and a revival beam-splitter with the
transmission coefficient $T$ are used. The ancillary beam carrying
correlated noise is expressed by $\hat{a}_{an}^{\prime }=\hat{a}_{an}+\sqrt{%
g_{b}}\hat{N}$, where $\hat{a}_{an}$\ is the ancillary beam and $g_{b}$
describes the magnitude of the correlated noise, which is an adjustable
parameter in experiments. The transmitted ($\hat{A}_{R}$) and reflected ($%
\hat{D}$) beams from the revival beam-splitter are $\hat{A}_{R}=\sqrt{\eta
_{A}T}\hat{A}+\sqrt{\left( 1-\eta _{A}\right) T}\hat{\upsilon}_{A}-\sqrt{1-T}%
\hat{a}_{an}+(\sqrt{\left( 1-\eta _{A}\right) g_{a}T}-\sqrt{\left(
1-T\right) g_{b}})\hat{N}$ and $\hat{D}=\sqrt{\eta _{A}\left( 1-T\right) }%
\hat{A}+(\sqrt{\left( 1-\eta _{A}\right) \left( 1-T\right) g_{a}}+\sqrt{%
Tg_{b}})\hat{N}+\sqrt{\left( 1-\eta _{A}\right) \left( 1-T\right) }%
\allowbreak \hat{\upsilon}_{A}+\sqrt{T}\hat{a}_{an}$, respectively. If the
values of $g_{b}$ and $T$ are chosen to satisfy the following expression 
\begin{equation}
\frac{g_{a}}{g_{b}}=\frac{1-T}{\left( 1-\eta _{A}\right) T},
\label{relation}
\end{equation}%
the noise on the output mode $\hat{A}_{R}$ will be removed totally. In this
case, the output mode becomes 
\begin{equation}
\hat{A}_{R}=\sqrt{\eta _{A}T}\hat{A}+\sqrt{\left( 1-\eta _{A}\right) T}\hat{%
\upsilon}_{A}-\sqrt{1-T}\hat{a}_{an},  \label{outmode}
\end{equation}%
which is immune from the excess noise. The excess noise is transferred onto
the reflected beam $\hat{D}$\ (which is abandoned) due to the use of the
beam-splitter. Thus the tripartite entanglement among $\hat{A}_{R}$, $\hat{B}
$\ and $\hat{C}$ is preserved by using a correlated noisy channel.

\section*{Experimental set-up}

The experimental set-up for distributing a mode of the tripartite entangled
optical beams over a noisy channel and the entanglement revival is shown in
Fig. 1(c). The nondegenerate optical parametric amplifiers (NOPAs) are
pumped by a common laser source, which is a continuous wave intracavity
frequency-doubled and frequency-stabilized Nd:YAP/LBO (Nd-doped YAlO$_{3}$
perorskite/lithium triborate) laser. Each of the NOPAs consists of an $%
\alpha $-cut type-II KTP crystal and a concave mirror. The front face of the
KTP crystal was coated to be used for the input coupler and the concave
mirror serves as the output coupler of the squeezed states. The
transmissions of the input coupler at 540 nm and 1080 nm are $99.8\%$ and $%
0.04\%$, respectively. The transmissions of the output coupler at 540 nm and
1080 nm are $0.5\%$ and $5.2\%$, respectively. A pair of $\hat{x}$-squeezed
and $\hat{p}$-squeezed states in two orthogonal polarizations are produced
by NOPA1 \cite{Yun2000}. The other $\hat{x}$-squeezed state is produced by
NOPA2. NOPAs are locked individually by using Pound-Drever-Hall method with
a phase modulation of 56 MHz on 1080 nm laser beam \cite{PDH}. Both NOPAs
are operated at deamplification condition, which corresponds to lock the
relative phase between the pump laser and the injected signal to (2n+1)$\pi $%
\ (n is the integer).

The tripartite entangled state of optical field at the sideband frequency of
2 MHz is obtained by combining three squeezed states of light with $-3.5$ dB
squeezing and $8.5$ dB anti-squeezing on two optical beam-splitters with
transmission coefficients $T_{1}=1/3$ and $T_{2}=1/2$, respectively [see
APPENDIX A]. For a real communication channel, the loss and noise are coming
from the environment as that shown in Fig. 1. The loss in the quantum
channel is mimicked by a beam-splitter composed by two polarization
beam-splitters (PBSs) and a half-wave plate placed between them. The noisy
channel is simulated by adding a Gaussian noise on a coherent beam with
electro-optic modulators (EOMs) and then the modulated beam is coupled with
the transmitted mode on a beam-splitter with transmission efficiency $\eta
_{A}$. When an ancillary beam with the correlated noise is mixed with the
transmitted mode on a beam-splitter of the transmission coefficient $T$, the
entanglement revival is completed. The covariance matrix of the output state
is measured by three homodyne detectors. The quantum efficiency of the
photodiodes used in the homodyne detectors are 95\%. The interference
efficiency on all beam-splitters are about 99\%.

\section*{The positive partial transposition criterion}

\begin{figure}[tbp]
\begin{center}
\includegraphics[width=90mm]{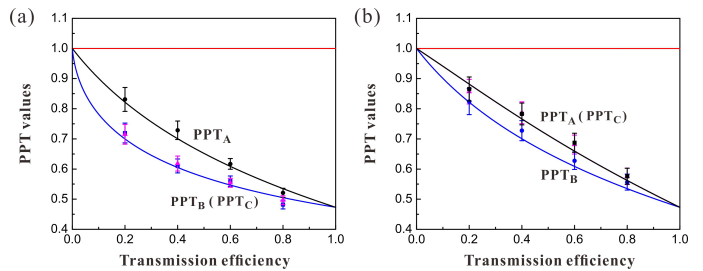}
\end{center}
\caption{\textbf{\ The entanglement in lossy channels.\ (a)}\ One optical
mode $\hat{A}$\ is distributed in the lossy channel ($\protect\eta _{A}\neq 1
$, $\protect\eta _{C}=1$).\ \textbf{(b)}\ Two optical modes $\hat{A}$\ and $%
\hat{C}$\ are distributed in the lossy channels with the same transmission
efficiency ($\protect\eta _{A}\neq 1$, $\protect\eta _{C}\neq 1$).\ PPT
values are all below the entanglement boundary (red lines), which means that
the tripartite entanglement is robust against loss in quantum channels. The
black, blue and pink dots represent the experimental data for different PPT
values, respectively. Error bars represent $\pm $\ one standard deviation
and are obtained based on the statistics of the measured noise variances.}
\end{figure}

The positive partial transposition (PPT) criterion \cite{Horodecki,Simon} is
a necessary and sufficient condition for judging the existence of quantum
entanglement among $N$\ Gaussian optical beams, when the state has the form
of bipartite splitting with only a single mode on one side like ($1|N-1$) 
\cite{Werner,Adesso}. We use the PPT criterion to verify the disentanglement
and the entanglement revival of the tripartite entangled states of light.
Based on above-mentioned expressions of output state, we obtain the
covariance matrix, which is given in APPENDIX B, and
calculate the symplectic eigenvalues. The positivity is checked by
evaluating the symplectic eigenvalues of the partially transposed matrix and
the state is separable if any of the symplectic eigenvalues is larger than
or equal to 1 \cite{Vollmer}.

At the level of quadrature operators, the partial transposition with respect
to mode $k$ ($k$ $=1,2,3$) corresponds to the change of sign of the phase
quadrature, i. e. $\hat{p}_{k}\longrightarrow -\hat{p}_{k}$. Symplectic
eigenvalues of covariance matrix are defined as the positive roots of the
polynomial $\left\vert \gamma ^{T(k)}-i\mu \Omega \right\vert =0$, where $%
\left\vert A\right\vert $ denotes the determinant of matrix \cite{Vollmer}. $%
\gamma ^{T(k)}=T_{k}\gamma T_{k}^{T}$ is the partially transposed matrix of
the quantum state, where $T_{k}$ is a diagonal matrix with all diagonal
elements equal to 1 except for $T_{2k,2k}=-1$, and 
\begin{equation}
\Omega =\oplus _{k=1}^{3}\left( 
\begin{array}{cc}
0 & 1 \\ 
-1 & 0%
\end{array}%
\right) .
\end{equation}%
We consider a bipartite splitting of a three-mode Gaussian state with
covariance matrix $\gamma $ such that one party holds mode $k$ and the other
party possesses the remained two modes. If the smallest symplectic
eigenvalue $\mu _{k}$ obtained from the polynomial is smaller than 1, the
state is inseparable with respect to the $k|ij$\ splitting.

\section*{Entanglement in lossy channels}

\begin{figure}[tbp]
\begin{center}
\includegraphics[width=90mm]{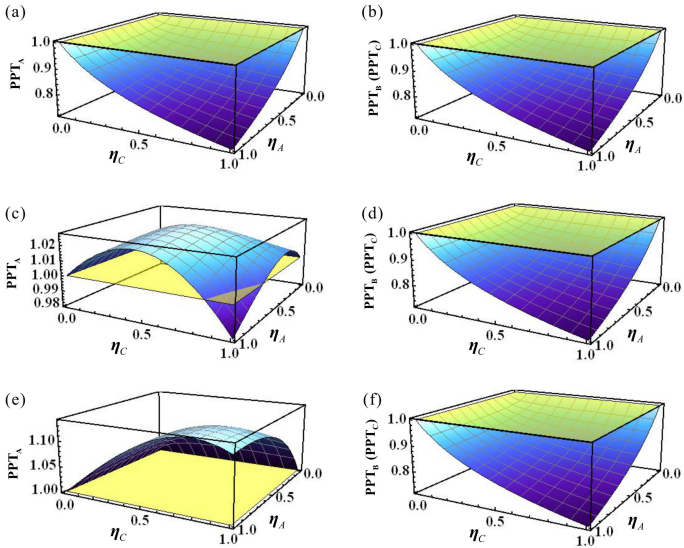}
\end{center}
\caption{\textbf{The entanglement properties of different asymmetric
tripartite states in two lossy channels.} (\textbf{a,b})\ PPT values of
fully robust entanglement against loss with\ $c_{x}/c=0.8$. (\textbf{c,d})\
PPT values of\ a one-mode fragile state in lossy channels for attenuations
with $c_{x}/c=0.5$. (\textbf{e,f}) a totally one-mode biseparable state in
lossy channels with $c_{x}/c=0.3$.}
\end{figure}
Figure 2 shows that the tripartite entanglement is robust against loss in
lossy quantum channels. The PPT values PPT$_{\text{A}}$, PPT$_{\text{B}}$
and PPT$_{\text{C}}$ represent the different splittings for the (A$|$BC), (B$%
|$AC) and (C$|$AB), respectively. The PPT values of distributing one optical
mode ($\hat{A}$) and two optical modes ($\hat{A}$\ and $\hat{C}$) in one and
two lossy channels (for simplification, we assume the losses in two quantum
channels are the same) are shown in Fig. 2(a) and Fig. 2(b), respectively.
The values of PPT$_{\text{B}}$\ and PPT$_{\text{C}}$\ in Fig. 2(a) [PPT$_{%
\text{A}}$\ and PPT$_{\text{C}}$\ in Fig. 2(b)] are the same because that
the optical modes of the tripartite entangled state are symmetric and modes $%
\hat{B}$\ and $\hat{C}$\ do not interact with the environment (modes $\hat{A}
$\ and $\hat{C}$\ are transmitted with the same transmission efficiency).
Comparing the corresponding PPT values in Fig. 2(a) and Fig. 2(b), we can
see that the tripartite entanglement is more robust if only a mode ($\hat{A}$%
) pass through the lossy channel than that both modes $\hat{A}$\ and $\hat{C}
$\ subject to the lossy channels, i. e. the degradation of entanglement in
the case of $\eta _{C}=1$\ is less than that in the case of $\eta _{A}\neq 1$%
\ and $\eta _{C}\neq 1$.\ In lossy channels, the tripartite entanglement
gradually decreases along with the degradation of the transmission
efficiency of quantum channel and finally tends to zero when the channel
efficiency equals to zero. This is different from the results in Ref. [25],
where the disentanglement of a tripartite entangled light beam is observed
over a lossy channel. We theoretically analyse the physical reason of the
difference based on the covariance matrix and discover that it is because
the tripartite entangled state prepared by us is a symmetric entangled
state, while the state in Ref. [25] is an asymmetric state since the effect
of the classical phonon noise, which are discussed detailedly at next
section.

\subsection*{Discussion on symmetric and asymmetric states}

We consider the case of the transmission in two lossy channels for
tripartite Gaussian symmetric and asymmetric optical states, respectively.
For convenience, the covariance matrix of the original tripartite Gaussian
state is written in terms of two-by-two submatrices as%
\begin{equation}
\sigma =\left[ 
\begin{array}{ccc}
\sigma _{A\text{ }} & \sigma _{AB} & \sigma _{AC} \\ 
\sigma _{AB}^{T} & \sigma _{B\text{ }} & \sigma _{BC} \\ 
\sigma _{AC}^{T} & \sigma _{BC}^{T} & \sigma _{C\text{ }}%
\end{array}%
\right] ,
\end{equation}%
in which each diagonal block $\sigma _{k\text{ }}$is the local covariance
matrix corresponding to the reduced state of mode\emph{\ }$k$\emph{\ }($%
k=A,B,C$)\emph{, }respectively, and the off-diagonal matrices $\sigma _{mn}$
are the intermodal correlations between subsystems $m$ and $n$. The detailed
expressions of $\sigma _{k\text{ }}$and $\sigma _{mn}$\ are given in APPENDIX B.

\begin{figure}[tbp]
\begin{center}
\includegraphics[width=90mm]{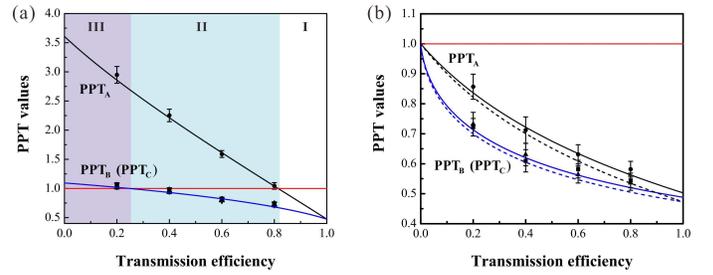}
\end{center}
\caption{\textbf{The disentanglement and entanglement revival in a noisy
channel.} (\textbf{a}) The PPT values for the transmission in a noisy
channel,\textbf{\ }where the variance of the excess noise is taken as five
times of shot noise level. The tripartite entangled state experiences
entanglement (I), one-mode biseparable (II) and fully disentanglement (III)
along with the decreasing of the channel efficiency. (\textbf{b}) The PPT
values after entanglement revival. Dash lines are the corresponding results
of the perfectly revival, the results are the same with the lines in Fig.
2(a) which are obtained before disentanglement. The black dots represent the
experimental data. Error bars represent $\pm $ one standard deviation and
are obtained based on the statistics of the measured noise variances.}
\end{figure}

\emph{Type I symmetric state.} If a quantum state has symmetric modes ($%
\sigma _{A\text{ }}=\sigma _{B\text{ }}=\sigma _{C}$) and balanced
correlations between subsystems $\sigma _{mn}$, i. e. the absolute values of
main diagonal elements in $\sigma _{mn}$ are the same, we say it is a
symmetric state, where the variances are $\bigtriangleup ^{2}\hat{x}%
_{A}=\bigtriangleup ^{2}\hat{x}_{B}=\bigtriangleup ^{2}\hat{x}_{C}=s$, $%
\bigtriangleup ^{2}\hat{p}_{A}=\bigtriangleup ^{2}\hat{p}_{B}=\bigtriangleup
^{2}\hat{p}_{C}=t$, $\left\langle \delta \hat{x}_{A}\delta \hat{x}%
_{B}\right\rangle =\left\langle \delta \hat{x}_{A}\delta \hat{x}%
_{C}\right\rangle =\left\langle \delta \hat{x}_{B}\delta \hat{x}%
_{C}\right\rangle =c$, $\left\langle \delta \hat{p}_{A}\delta \hat{p}%
_{B}\right\rangle =\left\langle \delta \hat{p}_{A}\delta \hat{p}%
_{C}\right\rangle =\left\langle \delta \hat{p}_{B}\delta \hat{p}%
_{C}\right\rangle =-c$, and $\left\langle \delta \hat{x}_{j}\delta \hat{x}%
_{j^{\prime }}\right\rangle =0$. The corresponding covariance matrix is 
\begin{equation}
\sigma _{I}=\left[ 
\begin{array}{cccccc}
s & 0 & c & 0 & c & 0 \\ 
0 & t & 0 & -c & 0 & -c \\ 
c & 0 & s & 0 & c & 0 \\ 
0 & -c & 0 & t & 0 & -c \\ 
c & 0 & c & 0 & s & 0 \\ 
0 & -c & 0 & -c & 0 & t%
\end{array}%
\right] .
\end{equation}%
This type of quantum state can be generated by the interference of three
squeezed states on beam-splitter network. The three squeezed states are
produced from three NOPAs operating below its oscillation threshold,
respectively \cite{Yun2000}. The experimental values of parameters c, s and
t in Eq. (6) are obtained by the covariance matrix of the entangled state
prepared by us [see Eq. (13) in APPENDIX B]. Apparently, the prepared state
is a symmetric state. According to the theoretical calculation result shown
in Fig. 2, in which all used parameters in the calculation are derived from
Eq. (13) and (14) in APPENDIX B, it is proved that the
symmetric quantum states are fully robust against losses in the two lossy
channels.

\emph{Type II asymmetric state.} The asymmetric quantum state has unbalanced
correlations between subsystems ($c_{x}\neq c$), whose covariance matrix is
given by

\begin{equation}
\sigma _{II}=\left[ 
\begin{array}{cccccc}
s & 0 & c_{x} & 0 & c_{x} & 0 \\ 
0 & t & 0 & -c & 0 & -c \\ 
c_{x} & 0 & s & 0 & c & 0 \\ 
0 & -c & 0 & t & 0 & -c \\ 
c_{x} & 0 & c & 0 & s & 0 \\ 
0 & -c & 0 & -c & 0 & t%
\end{array}%
\right] .  \label{asym}
\end{equation}

The entanglement property of the asymmetric state is shown in Fig. 3 [all
parameters used in the calculation are also taken from Eq. (13) and (14) in
APPENDIX B]. The mathematic operation of modifying a parameter, $c_{x}$, and
keeping other elements in the covariance matrix unchanging is equivalent to
physically add an uncorrelated noise into the generation system of the
tripartite entangled state \cite{Barbosa2}. For example, the phonon noise in
the prepared three-color entangled state in Ref. [25], which is produced by
a non-degenerated optical parametric oscillator (NOPO) operating above its
oscillation threshold, is a type of classical and uncorrelated noises. In an
above-threshold NOPO, the effect of the uncorrelated phonon noise is large
and thus the produced three-color entangled state is asymmetric \cite{Coelho}%
. In Eq. (\ref{asym}), we assume that the different amounts of the
uncorrelated noise exist in mode $\hat{A}$, thus $c_{x}\ $in both $\sigma
_{AB}$ and $\sigma _{AC}$ is changed simultaneously, while $\sigma _{BC}$ is
unchangded because it is not related to mode $\hat{A}$. Fig. 3(a) and 3(b)
are the PPT values corresponding to $c_{x}/c=0.8$, which show that the
disentanglement never happens, thus the original state is robust against
losses. Actually, all asymmetric states correspond to $c_{x}/c>0.8$\ are
robust against losses according to our calculation. Fig. 3(c) and 3(d) are
the PPT values corresponding to $c_{x}/c=0.5$, which show that the
disentanglement happens for A$|$BC during decreasing the transmission
efficiencies, thus the original state is an one-mode fragile state for the
attenuations. From Fig. 3(e) and 3(f) we can see that the state is a totally
one-mode biseparable state when $c_{x}/c=0.3$. Thus the original asymmetric
tripartite state evolutes from a robust state to a fully one-mode
biseparable state with the decrease of the $c_{x}$\ value. We should emphase
that that the asymmetric property of quantum states does not certainly
result in the disentanglement, and only when the uncorrelated noise in the
quantum state is large enough the disentanglement occurs. To the physical
reason, the correlation among subsystems in a symmetric state is balanced,
i. e. there is no uncorrelated noise in the quantum state. Thus all
entangled modes for the symmetric state are equivalent and the entanglement
is reduced gradually and continually in a lossy channel. However, for an
asymmetric state, the correlation among subsystems is unbalanced, i. e.
uncorrelated noises are added into the quantum state, which lead to
disentanglement of the asymmetric tripartite state.

\begin{figure}[tbp]
\begin{center}
\includegraphics[width=90mm]{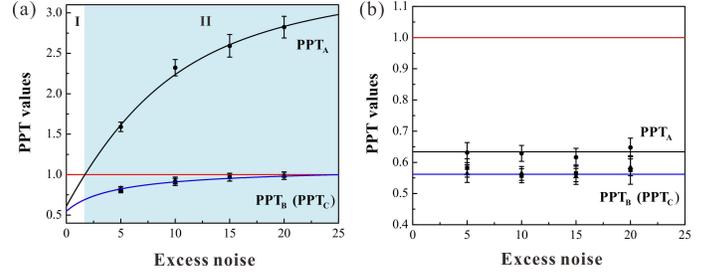}
\end{center}
\caption{\textbf{The disentanglement and revival of entanglement at
different noise levels (in the unit of shot noise level).} (\textbf{a})
Disentanglement at different excess noise levels. (\textbf{b}) The PPT
values after entanglement revival, \ and it is independent on the excess
noise. The black dots represent the experimental data. Error bars represent $%
\pm $ one standard deviation and are obtained based on the statistics of the
measured noise variances.}
\end{figure}

\section*{Entanglement in noisy channel}

When there is the excess noise in the quantum channel, the disentanglement
is observed as shown in Fig. 4, where the variance of the excess noise is
taken as five times of shot noise level and $g_{a}=1$. The values of PPT$_{%
\text{B}}$ and PPT$_{\text{C}}$ are the same because the tripartite
entangled state is symmetric and both modes $\hat{B}$ and $\hat{C}$ are
retained in a node. In this case, we can assume that modes $\hat{B}$ and $%
\hat{C}$ have no interaction with the environment. Entanglement survives
when the transmission efficiency satisfies $0.81<\eta \leq 1$ [region I in
Fig. 4(a)]. When $0.24<\eta \leq 0.81,$\ PPT$_{\text{A}}$ is above 1 while
PPT$_{\text{B}}$ and PPT$_{\text{C}}$ are below 1, the state is
corresponding to a one-mode biseparable state [region II in Fig. 4(a)],
which means that mode $\hat{A}$ is separated from modes $\hat{B}$ and $\hat{C%
}$. When the transmission efficiency is $\eta \leq 0.24$, fully
disentanglement is observed [region III in Fig. 4(a)], which will result in
that the quantum communication between any two users is not possible to be
implemented.

There are two adjustable parameters in the entanglement revival procedure, $%
g_{b}$ and $T$. For different channel efficiency, we may fix one of them and
adjust the other one to recover the entanglement according to Eq. (\ref%
{relation}). In the experiment, we chose to fix $T$ and adjust $g_{b}$ for
different transmission efficiencies of the quantum channel. Generally, $T$
can not be taken too small because it corresponds to add a linear loss on
the transmitted mode, which will degrade the tripartite entanglement. The
solid and dash lines in Fig. 4(b) represent the revived entanglement in our
experiment with $T=90\%$ and in the ideal case with the perfectly revival,
respectively. The imperfect transmission efficiency of the revival
beam-splitter leads to the small difference between the two results. The
parameters $g_{a}/g_{b}$ are chosen to be 0.14, 0.18, 0.28 and 0.56 for
channel efficiencies of 0.2, 0.4, 0.6 and 0.8, respectively. Fig. 4(b) shows
that the entanglement is revived after the revival operation.

The dependence of entanglement on the excess noise is shown in Fig. 5, where
the transmission efficiency is chosen to be 0.6, $g_{a}/g_{b}=0.28$, and $%
T=90\%$. We can see that the boundary of disentanglement depends on the
excess noise level.\ When the variance of the excess noise is lower than 2.2
times of shot noise level for the transmission efficiency of $\eta =0.6$,
the entanglement can be survived in a noisy channel. When the variance of
the excess noise is higher than 2.2 times of shot noise level for the
transmission efficiency $\eta =0.6$, the disentanglement happens and the
state is reduced to an one-mode biseparable state [region II in Fig. 5(a)].
After the revival operation, the entanglement is recovered and it is
independent on the excess noise as that indicated by Eq. (3).

\section*{Discussion}

In summary, we deeply investigate the different effects of the lossy channel
and the noisy channel on the tripartite entangled state of light when it is
distributed in a quantum network and point out that the entangled optical
beams with the symmetric structure is more robust to the channel losses than
that with the asymmetric structure. From the experimental and the calculated
results we can conclude that the symmetry is a sufficient but not necessary
condition for robustness of tripartite entanglement in a lossy channel.
While for the asymmetric state, the robustness of the state depends on the
correlation between subsystems.

Disentanglement is observed when the excess noise exists in the quantum
channel. The symmetric tripartite entangled state is fragile against the
excess noise, while it is robust against the pure loss. By creating a
correlated noisy channel (non-Markovian environment), entanglement of the
tripartite entangled state is preserved, thus disentanglement can be avoided
with the correlated noisy channel. But when the revived state passed through
a noisy channel again the disentanglement will possibly appear again.

The influence of loss and noise in quantum channel to the entanglement
depends on the strength of them. The more the strength is, the large the
influence is. The correlated noisy channel used in our experiment can only
remove the effect of noise in the quantum channel, while the influence of
loss can not be eliminated by the scheme. Fortunately, another technology,
the noiseless linear amplification, can be used to eliminate the effect of
loss on entanglement \cite{Ralph,Xiang,Helen,Alex}. Due to that the
disentanglement induced by environment is a specific phenomenon among
correlated quantum systems, which is never observed in the studies of
dissipation effects for classical systems, the presented results are
significant to understand the dynamic behavior of the interaction of
entangled states and different environment. Besides, our investigation also
provides concrete references for establishing quantum network with
multipartite entangled states of light.

\textbf{Acknowledgements }This research was supported by NSFC (Grant Nos.
11522433, 61475092) and OIT (2013805). X. Su and C. Xie conceived the
original idea. X. Deng and X. Su designed the experiment. X. Deng, C. Tian
and X. Su constructed and performed the experiment. X. Deng and C. Tian
accomplished theoretical calculation. X. Deng, C. Tian and X. Su,
accomplished the data analyses. X. Su and C. Xie wrote the paper.

\section*{APPENDIX}

\subsection{Preparation of the tripartite entangled state}

\begin{figure}[tbp]
\begin{center}
\includegraphics[width=85mm]{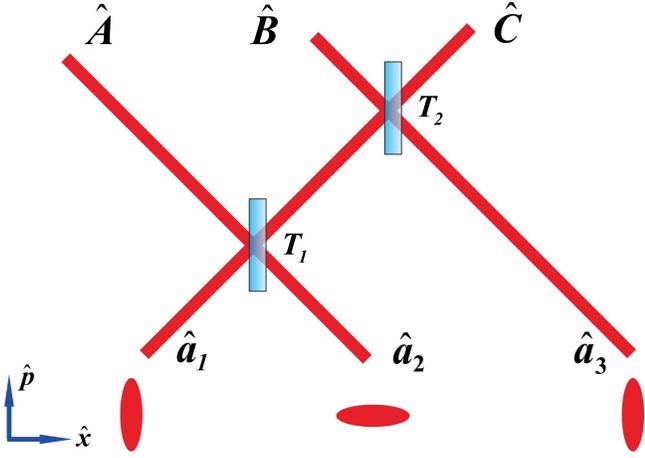}
\end{center}
\caption{The schematic of generation system of the tripartite entangled
state.}
\end{figure}

The tripartite entangled state used in the experiment is a continuous
variable tripartite Greenberger-Horne-Zeilinger (GHZ) state of optical field 
\cite{Loock} which is prepared by coupling a phase-squeezed state ($\hat{a}%
_{2}$) of light and two amplitude-squeezed states of light ($\hat{a}_{1}$
and $\hat{a}_{3}$) on an optical beam-splitter network,\ which consists of
two optical beam-splitters with transmittance of $T_{1}=1/3$ and $T_{2}=1/2$%
, respectively, as shown in Fig. 6. Three input squeezed states are
expressed by

\begin{align}
\hat{a}_{1}& =e^{-r_{1}}\hat{x}_{1}^{(0)}+ie^{r_{1}}\hat{p}_{1}^{(0)}, 
\notag \\
\hat{a}_{2}& =e^{r_{2}}\hat{x}_{2}^{(0)}+ie^{-r_{2}}\hat{p}_{2}^{(0)}, 
\notag \\
\hat{a}_{3}& =e^{-r_{3}}\hat{x}_{3}^{(0)}+ie^{r_{3}}\hat{p}_{3}^{\left(
0\right) },
\end{align}%
where $r_{i}$ ($i=1,2,3$) is the squeezing parameter, $\hat{x}=\hat{a}+\hat{a%
}^{\dag }$ and $\hat{p}=(\hat{a}-\hat{a}^{\dag })/i$ are the amplitude and
phase quadratures of an optical field $\hat{a}$, respectively, and the
superscript of the amplitude and phase quadratures represent the vacuum
state. The transformation matrix of the beam-splitter network is given by

\begin{equation}
U=\left[ 
\begin{array}{ccc}
\sqrt{\frac{2}{3}} & \sqrt{\frac{1}{3}} & 0 \\ 
-\sqrt{\frac{1}{6}} & \sqrt{\frac{1}{3}} & \sqrt{\frac{1}{2}} \\ 
-\sqrt{\frac{1}{6}} & \sqrt{\frac{1}{3}} & -\sqrt{\frac{1}{2}}%
\end{array}%
\right] ,
\end{equation}%
the unitary matrix can be decomposed into a beam-splitter network $%
U=B_{23}^{+}(T_{2})I_{2}(-1)B_{12}^{+}(T_{1}),$ where $B_{kl}^{+}(T_{j})$
stands for the linearly optical transformation on $j$th beam-splitter with
transmission of $T_{j}$ ($j=1,2$), where $\left( B_{kl}^{+}\right) _{kk}=%
\sqrt{1-T},\left( B_{kl}^{+}\right) _{kl}=\left( B_{kl}^{+}\right) _{lk}=%
\sqrt{T},\left( B_{kl}^{+}\right) _{ll}=-\sqrt{1-T},$ are matrix elements of
the beam-splitter. $I_{k}(-1)=e^{i\pi }$ corresponds to a $180%
{{}^\circ}%
$ rotation in phase space. The output modes from the optical beam-splitter
network are expressed by 
\begin{align}
\hat{A}& =\sqrt{\frac{2}{3}}\hat{a}_{1}+\sqrt{\frac{1}{3}}\hat{a}_{2}, 
\notag \\
\hat{B}& =-\sqrt{\frac{1}{6}}\hat{a}_{1}+\sqrt{\frac{1}{3}}\hat{a}_{2}+\sqrt{%
\frac{1}{2}}\hat{a}_{3},  \notag \\
\hat{C}& =-\sqrt{\frac{1}{6}}\hat{a}_{1}+\sqrt{\frac{1}{3}}\hat{a}_{2}-\sqrt{%
\frac{1}{2}}\hat{a}_{3},
\end{align}%
respectively. Here, we have assumed that three squeezed states have the
identical squeezing parameter ($r_{1}=r_{2}=r_{3}$). In experiments, the
requirement is easily achieved by adjusting the two nondegenerate optical
parametric amplifiers (NOPAs) to operate at totally same conditions. For our
experimental system, we have $r=0.4$.

\subsection{Covariance matrix of the tripartite optical beams}

Gaussian state is the state with Gaussian characteristic functions and
quasi-probability distributions on the multi-mode quantum phase space, which
can be completely characterized by a covariance matrix. The elements of the
tripartite covariance matrix are $\sigma _{ij}=Cov\left( \hat{R}_{i},\hat{R}%
_{j}\right) =\frac{1}{2}\left\langle \hat{R}_{i}\hat{R}_{j}+\hat{R}_{j}\hat{R%
}_{i}\right\rangle -\left\langle \hat{R}_{i}\right\rangle \left\langle \hat{R%
}_{j}\right\rangle $, $i,j=1,2,\ldots ,6$, where $\hat{R}=(\hat{x}_{A},\hat{p%
}_{A},\hat{x}_{B},\hat{p}_{B},\hat{x}_{C},\hat{p}_{C})^{T}$ is a vector
composed by the amplitude and phase quadratures of tripartite optical beams 
\cite{Adesso2}. Thus covariance matrix of tripartite optical beams can be
partially expressed as (the cross correlations between different quadratures
are taken as $0$)%
\begin{align}
\sigma _{A\text{ }}& =\left[ 
\begin{array}{cc}
\bigtriangleup ^{2}\hat{x}_{A} & 0 \\ 
0 & \bigtriangleup ^{2}\hat{p}_{A}%
\end{array}%
\right] ,  \notag \\
\sigma _{B\text{ }}& =\left[ 
\begin{array}{cc}
\bigtriangleup ^{2}\hat{x}_{B} & 0 \\ 
0 & \bigtriangleup ^{2}\hat{p}_{B}%
\end{array}%
\right] ,  \notag \\
\sigma _{C\text{ }}& =\left[ 
\begin{array}{cc}
\bigtriangleup ^{2}\hat{x}_{C} & 0 \\ 
0 & \bigtriangleup ^{2}\hat{p}_{C}%
\end{array}%
\right] ,  \notag \\
\sigma _{AB\text{ }}& =\left[ 
\begin{array}{cc}
Cov\left( \hat{x}_{A},\hat{x}_{B}\right) & 0 \\ 
0 & Cov\left( \hat{p}_{A},\hat{p}_{B}\right)%
\end{array}%
\right] ,  \notag \\
\sigma _{AC\text{ }}& =\left[ 
\begin{array}{cc}
Cov\left( \hat{x}_{A},\hat{x}_{C}\right) & 0 \\ 
0 & Cov\left( \hat{p}_{A},\hat{p}_{C}\right)%
\end{array}%
\right] ,  \notag \\
\sigma _{BC\text{ }}& =\left[ 
\begin{array}{cc}
Cov\left( \hat{x}_{B},\hat{x}_{C}\right) & 0 \\ 
0 & Cov\left( \hat{p}_{B},\hat{p}_{C}\right)%
\end{array}%
\right] .
\end{align}

To partially reconstruct all relevant entries of its associated covariance
matrix we perform 12 different measurements on the output optical modes.
These measurements include the amplitude and phase quadratures of the output
optical modes, and the cross correlations $\Delta ^{2}\left( \hat{x}_{A}-%
\hat{x}_{B}\right) $, $\Delta ^{2}\left( \hat{x}_{A}-\hat{x}_{C}\right) $, $%
\Delta ^{2}\left( \hat{x}_{B}-\hat{x}_{C}\right) $, $\Delta ^{2}\left( \hat{p%
}_{A}+\hat{p}_{B}\right) $, $\Delta ^{2}\left( \hat{p}_{A}+\hat{p}%
_{C}\right) $ and $\Delta ^{2}\left( \hat{p}_{B}+\hat{p}_{C}\right) $. The
covariance elements are calculated via the identities \cite{Steinlechner}%
\begin{align}
Cov\left( \hat{R}_{i},\hat{R}_{j}\right) & =\frac{1}{2}\left[ \Delta
^{2}\left( \hat{R}_{i}+\hat{R}_{j}\right) -\Delta ^{2}\hat{R}_{i}-\Delta ^{2}%
\hat{R}_{j}\right] ,  \notag \\
Cov\left( \hat{R}_{i},\hat{R}_{j}\right) & =-\frac{1}{2}\left[ \Delta
^{2}\left( \hat{R}_{i}-\hat{R}_{j}\right) -\Delta ^{2}\hat{R}_{i}-\Delta ^{2}%
\hat{R}_{j}\right] .
\end{align}%
In the experiment, we obtain all the covariance matrices of every quantum
state actually, and then calculate the PPT value to verify whether the
quantum states are entangled or not.

The partially reconstructed covariance matrix of the prepared tripartite
entangled optical field is 
\begin{equation}
\sigma =\left( 
\begin{array}{cccccc}
2.76 & 0 & 2.32 & 0 & 2.27 & 0 \\ 
0 & 5.05 & 0 & -2.23 & 0 & -2.27 \\ 
2.32 & 0 & 2.78 & 0 & 2.29 & 0 \\ 
0 & -2.23 & 0 & 4.81 & 0 & -2.14 \\ 
2.27 & 0 & 2.29 & 0 & 2.69 & 0 \\ 
0 & -2.27 & 0 & -2.14 & 0 & 4.80%
\end{array}%
\right) .
\end{equation}%
The entanglement among the prepared tripartite state is evaluated by PPT
criterion and we have PPT values PPT$_{\text{A}}=0.48$, PPT$_{\text{B}}=0.47$%
, PPT$_{\text{C}}=0.48$, respectively.

For the tripartite state distributed over lossy (noisy) channels in the
experiment we have 
\begin{align}
\sigma _{A\text{ }}^{\prime }& =\eta _{A}\sigma _{A\text{ }}+\left( 1-\eta
_{A}\right) \left( g_{a}N+1\right) I,  \notag \\
\sigma _{B\text{ }}^{\prime }& =\sigma _{B\text{ }},  \notag \\
\sigma _{C\text{ }}^{\prime }& =\eta _{C}\sigma _{C\text{ }}+\left( 1-\eta
_{C}\right) I,  \notag \\
\sigma _{AB}^{\prime }& =\sqrt{\eta _{A}}\sigma _{AB},  \notag \\
\sigma _{AC}^{\prime }& =\sqrt{\eta _{A}\eta _{C}}\sigma _{AC},  \notag \\
\sigma _{BC}^{\prime }& =\sqrt{\eta _{C}}\sigma _{BC},
\end{align}%
\newline
where $\eta _{A}$ and $\eta _{C}$ are the transmission efficiencies of
optical modes $\hat{A}$ and $\hat{C}$ respectively, $N$ and $g_{a}$
represent the excess Gaussian noise in the channel and the magnitude of
noise respectively. Different combinations of the parameters correspond to
different transmission scenarios. When $\eta _{A}\neq 1$, $\eta _{C}\neq 1$
and $g_{a}N=0$, the optical modes $\hat{A}$ and $\hat{C}$ are distributed
over two lossy channels, respectively, which corresponds to the scenario
shown in Fig. 1(a). If $\eta _{A}\neq 1$, $\eta _{C}=1$ and $g_{a}N=0$, only
mode $\hat{A}$ is distributed over a lossy channel, while modes $\hat{B}$
and $\hat{C}$ are placed within a node without any channel between them.
When the optical mode $\hat{A}$ is distributed over a noisy channel, the
scenario is shown in Fig. 1(b). In this case we have $\eta _{A}\neq 1$, $%
\eta _{C}=1$ and $g_{a}N\neq 0$. After the entanglement revival operation,
the elements of the covariance matrix for the output state are%
\begin{align}
\sigma _{A\text{ }}^{\prime \prime }& =\eta _{A}T\sigma _{A\text{ }}+\left(
1-\eta _{A}T\right) I,  \notag \\
\sigma _{B\text{ }}^{\prime \prime }& =\sigma _{B\text{ }},  \notag \\
\sigma _{C\text{ }}^{\prime \prime }& =\sigma _{C\text{ }},  \notag \\
\sigma _{AB}^{\prime \prime }& =\sqrt{\eta _{A}T}\sigma _{AB},  \notag \\
\sigma _{AC}^{\prime \prime }& =\sqrt{\eta _{A}T}\sigma _{AC},  \notag \\
\sigma _{BC}^{\prime \prime }& =\sigma _{BC},
\end{align}%
respectively.

\newpage 


\begin{thebibliography}{99}
\bibitem{Nielson} Nielson, M. A. \& Chuang, I. L. \textit{Quantum
Computation And Quantum Information}. (Cambridge University, Cambridge,
2000).

\bibitem{Lampingk} Ralph, T. C. \& Lam, P. K. A bright future for quantum
communications. \textit{Nat. Photon.} \textbf{3,} 671-673 (2009).

\bibitem{Braunstein} Braunstein, S. L. \& van Loock, P. Quantum information
with continuous variables. \textit{Rev. Mod. Phys.} \textbf{77,} 513-577
(2005).

\bibitem{Weedbrook} Weedbrook, C. \textit{et al.} Gaussian quantum
information. \textit{Rev. Mod. Phys.} \textbf{84,} 621-669 (2012).

\bibitem{Raussendorf2001} Raussendorf, R. \& Briegel, H. J. A one-way
quantum computer. \textit{Phys. Rev. Lett.} \textbf{86,} 5188-5191 (2001).

\bibitem{Walther2005} Walther, P. \textit{et al.} Experimental one-way
quantum computing. \textit{Nature} \textbf{434,} 169-176 (2005).

\bibitem{Ukai2011} Ukai, R. \textit{et al.} Demonstration of unconditional
one-way quantum computations for continuous variables. \textit{Phys. Rev.
Lett.} \textbf{106,} 240504 (2011).

\bibitem{Su} Su, X. \textit{et al.} Gate sequence for continuous variable
one-way quantum computation. \textit{Nat. Commun}. \textbf{4,} 2828 (2013).

\bibitem{Loock} van Loock, P. \& Braunstein, S. L. Multipartite entanglement
for contonuous variables: a quantum teleportation network. \textit{Phys.
Rev. Lett.} \textbf{84,} 3482-3485 (2000).

\bibitem{Hide} Yonezawa, H., Aoki, T. \& Furusawa, A. Demonstration of a
quantum teleportation network for continuous variables. \textit{Nature} 
\textbf{431,} 430-433 (2004).

\bibitem{Jing} Jing, J. \textit{et al.} Experimental demonstration of
tripartite entanglement and controlled dense coding for continuous
variables. \textit{Phys. Rev. Lett.} \textbf{90,} 167903 (2003).

\bibitem{Jona} Roslund, J., de Ara\'{u}jo, R. M., Jiang, S., Fabre, C. \&
Treps, N. Wavelength-multiplexed quantum networks with ultrafast frequency
combs. \textit{Nat. Photon.} \textbf{8,} 109-112 (2014).

\bibitem{Pan} Pan, J. W., Simon, C., Brukner, \v{C}. \& Zeilinger, A.
Entanglement purification for quantum communication. \textit{Nature }\textbf{%
410,} 1067 (2001).

\bibitem{Duan} Duan, L. M., Giedke, G., Cirac, J. I. \& Zoller, P.
Entanglement purification of Gaussian continuous variable quantum states. 
\textit{Phys. Rev. Lett.} \textbf{84,} 4002 (2000).

\bibitem{Hage} Hage, B. \textit{et al. }Preparation of distilled and
purified continuous-variable entangled states. \textit{Nat. Phys. }\textbf{%
4, }915 (2008).

\bibitem{Dong} Dong, R.\ \textit{et al. }Experimental entanglement
distillation of mesoscopic quantum states. \textit{Nat. Phys. }\textbf{4,}
919 (2008).

\bibitem{Yu} Yu, T. \& Eberly, J. H. Sudden death of entanglement. \textit{%
Science} \textbf{323,} 598-601 (2009).

\bibitem{Almeida} Almeida, M. P. \textit{et al.} Environment-induced sudden
death of entanglement. \textit{Science} \textbf{316,} 579-582 (2007).

\bibitem{Lopez} L\'{o}pez, C. E.,\textit{\ }Romero, G., Lastra, F., Solano,
E. \& Retamal, J. C.\textit{\ }Sudden birth versus sudden death of
entanglement in multipartite systems. \textit{Phys. Rev. Lett.} \textbf{101,}
080503 (2008).

\bibitem{Aolita} Aolita, L., Chaves, R., Cavalcanti, D., Ac\'{\i}n, A. \&
Davidovich, L. Scaling laws for the decay of multiqubit entanglement. 
\textit{Phys. Rev. Lett.} \textbf{100,} 080501 (2008).

\bibitem{Xu} Xu, J. S. \textit{et al. }Experimental demonstration of
photonic entanglement collapse and revival. \textit{Phys. Rev. Lett.} 
\textbf{104,} 100502 (2010).

\bibitem{Kim} Kim, Y. S., Lee, J. C., Kwon, O. \& Kim Y. H. Protecting
entanglement from decoherence using weak measurement and quantum measurement
reversal. \textit{Nat. Phys.} \textbf{8,} 117-120 (2012).

\bibitem{Yamamoto} Yamamoto, N., Nurdin, H. I., James, M. R. \& Petersen, I.
R. Avoiding entanglement sudden death via measurement feedback control in a
quantum network. \textit{Phys. Rev. A\ }\textbf{78,} 042339 (2008).

\bibitem{Barbosa1} Barbosa, F. A. S. \textit{et al.} Robustness of bipartite
Gaussian entangled beams propagating in lossy channels. \textit{Nat. Photon.}
\textbf{4,} 858-861 (2010).

\bibitem{Coelho} Coelho, A. S. \textit{et al.} Three-color entanglement. 
\textit{Science} \textbf{326,} 823-826 (2009).

\bibitem{Kret} Kretschmann, D. \& Werner, R. F. Quantum channels with
memory. \textit{Phys. Rev. A} \textbf{72,} 062323 (2005).

\bibitem{Corney} Corney, J. F. \textit{et al.} Many-Body Quantum dynamics of
polarization squeezing in optical fibers. \textit{Phys. Rev. Lett.}\textbf{\
97,} 023606 (2006).

\bibitem{Andersen} Lassen, M., Berni, A., Madsen, L. S., Filip, R. \&
Andersen, U. L. Gaussian error correction of quantum states in a correlated
noisy channel. \textit{Phys. Rev. Lett.} \textbf{111,} 180502 (2013).

\bibitem{Deng} Deng, X. \textit{et al. }Disappearance and revival of
squeezing in quantum communication with squeezed state over a noisy channel. 
\textit{Appl. Phys. Lett.} \textbf{108,} 081105 (2016).

\bibitem{Yun2000} Zhang, Y. \textit{et al.} Experimental generation of
bright two-mode quadrature squeezed light from a narrow-band nondegenerate
optical parametric amplifier. \textit{Phys. Rev. A} \textbf{62,} 023813
(2000).

\bibitem{PDH} R. W. P. Drever \textit{et al.}, Laser phase and frequency
stabilization using an optical resonator. \textit{Appl. Phys. B} \textbf{31,}
97 (1983).

\bibitem{Horodecki} Horodecki, M., Horodecki, P., \& Horodecki, R.
Separability of mixed states: necessary and sufficient conditions. \textit{%
Phys. Lett. A} \textbf{223,} 1-8 (1996).

\bibitem{Simon} Simon, R. Peres-Horodecki separability criterion for
continuous variable systems. \textit{Phys. Rev. Lett.} \textbf{84,}
2726-2729 (2000).

\bibitem{Werner} Werner, R. F. \& Wolf, M. M. Bound entangled Gaussian
states. \textit{Phys. Rev. Lett.} \textbf{86,} 3658-3661 (2001).

\bibitem{Adesso} Adesso, G., Serafini, A. \& Illuminati, F. Multipartite
entanglement in three-mode Gaussian states of continuous-variable systems:
Quantification, sharing structure, and decoherence. \textit{Phys. Rev. A\ }%
\textbf{73,} 032345(2006).

\bibitem{Vollmer} Vollmer, C. E. \textit{et al.} Experimental entanglement
distribution by separable states. \textit{Phys. Rev. Lett.} \textbf{111,}
230505 (2013).

\bibitem{Barbosa2} Barbosa, F. A. S. \textit{et al}. Disentanglement in
bipartite continuous-variable systems. \textit{Phys. Rev. A} \textbf{84,}
052330 (2011).

\bibitem{Ralph} Ralph, T. C. Quantum error correction of continuous-variable
states against gaussian noise. \textit{Phys. Rev. A} \textbf{84,} 022339
(2011).

\bibitem{Xiang} Xiang, G. Y., Ralph, T. C., Lund, A. P., Walk, N. \& Pryde,
G. J. Heralded noiseless linear amplification and distillation of
entanglement. \textit{Nature Photon.} \textbf{4,} 316--319 (2010).

\bibitem{Helen} Chrzanowski, H. M. \textit{et al}. Measurement-based
noiseless linear amplification for quantum communication. \textit{Nature
Photon.} \textbf{8,} 333-338 (2014).

\bibitem{Alex} Ulanov, A. E. \textit{et al}. Undoing the effect of loss on
quantum entanglement. \textit{Nature Photon.} \textbf{9,} 764-768 (2015).

\bibitem{Adesso2} Adesso, G. \& Illuminati, F. \textit{J. Phys. A: Math.
Theor.} \textbf{40,} 7821 (2007).

\bibitem{Steinlechner} Steinlechner, S., Bauchrowitz, J., Eberle, T. \&
Schnabel, R. \textit{Phys. Rev. A} \textbf{87,} 022104 (2013).
\end{thebibliography}
\end{document}